\documentclass[11pt]{article}
\usepackage[margin=1in]{geometry}
\usepackage{amsmath,amssymb,amsthm}
\usepackage{tikz}
\usetikzlibrary{arrows.meta}
\usepackage[hidelinks]{hyperref}
\hypersetup{
  pdftitle={Computing EF1 Allocations with Few Value Queries},
  pdfauthor={Max Dupré la Tour and Ayumi Igarashi}
}
\newtheorem{theorem}{Theorem}[section]
\newtheorem{lemma}[theorem]{Lemma}
\theoremstyle{definition}
\newtheorem{algorithm}{Algorithm}
\theoremstyle{plain}

\title{Computing EF1 Allocations with Few Value Queries}
\author{Max Dupré la Tour \qquad Ayumi Igarashi\\[0.5em]
\normalsize RIKEN Center for Advanced Intelligence Project\\
\normalsize University of Tokyo}
\date{}

\begin{document}
\maketitle

\begin{abstract}
We give a deterministic polynomial-time algorithm that allocates $m$ indivisible goods
among $n$ agents with nonnegative additive valuations using
$O(n^5\log^2 m)$ value queries and guarantees envy-freeness up to one
good (EF1). Concurrent work of Lin, Liu, Tao, and Zhou achieves
$O(n^8\log m)$ comparison queries, attaining the optimal $O(\log m)$
dependence on $m$ for fixed $n$. Each comparison query can be simulated
using two value queries, so their bound also holds for value queries.
Our contributions are a much simpler algorithm and proofs,
and polynomial-time computation.
\end{abstract}

\section{Introduction}
\label{sec:introduction}

We study how to allocate indivisible goods fairly among agents with
different preferences. A central fairness requirement is \emph{envy-freeness}
(EF)~\cite{Varian74}, under which no agent prefers another agent's
bundle to her own. However, an envy-free allocation need not exist:
if two agents must divide a single good that both value positively,
whoever receives it is envied by the other agent.

A standard relaxation is \emph{envy-freeness up to one good}
(EF1)~\cite{Budish11}: for every pair of agents, removing at most
one good from the second agent's bundle must eliminate the first
agent's envy toward that bundle. For monotone valuations, the classical
envy-cycle elimination algorithm of Lipton et al.~\cite{LMMS04}
guarantees an EF1 allocation.

Finding a fair allocation requires information about the agents'
preferences. In the \emph{value-query model}, an algorithm asks an
agent for the exact numerical value of a specified bundle.
The query complexity counts these requests, separately from the
computation performed by the algorithm. With $n$ agents and $m$ goods,
envy-cycle elimination admits an implementation in polynomial time using
$O(mn)$ value queries~\cite{OPS21}. This bound applies to arbitrary monotone
valuations.

We focus on the regime in which the number of agents is small while
the number of goods is large. Asking each agent to report a value for
every good can be time-consuming. This concern persists even for
nonnegative additive valuations,
where a bundle's value is the sum of its goods' values and the entire
instance can be specified by $mn$ numbers. Value queries permit the
algorithm to evaluate large bundles directly, raising the possibility
of computing an EF1 allocation without learning all singleton values.

Oh, Procaccia, and Suksompong~\cite{OPS21} established that
$O(\log m)$ value queries suffice for two or three agents with additive
valuations. These results motivate studying the dependence on $m$ for
an arbitrary fixed number of agents. In this paper, we develop a simple
polynomial-time algorithm for additive valuations based on the
envy-cycle elimination algorithm.

\subsection{Our results}

\begin{theorem}[Value queries]
\label{thm:values}
There is a deterministic algorithm that, for $n\ge2$ agents with
nonnegative additive valuations over $m>n$ goods, finds a complete
EF1 allocation using $O(n^5\log^2 m)$ value queries.
It runs in time polynomial in $n,m$ in the exact real-arithmetic model.
\end{theorem}

For fixed $n$, the query bound is $O(\log^2 m)$. Our running-time
bound uses unit-cost exact arithmetic and comparisons. It excludes
oracle evaluation time.

\subsection{Concurrent work}
\label{sec:concurrent}

In concurrent work, Lin, Liu, Tao, and Zhou~\cite{LLTZ26} give an
$O(n^8\log m)$ comparison-query algorithm for additive goods and chores.
A comparison query asks an agent which of two bundles she weakly prefers.
Their dependence on $m$ is better than ours and is tight for fixed $n$,
matching the known $\Omega(\log m)$ lower bound for deterministic
algorithms~\cite{OPS21}. Their bound also yields
$O(n^8\log m)$ value queries by simulating each comparison with two
value queries. Thus our contribution is not a sharper dependence on $m$.

Our contributions are a simpler proof and polynomial running time.
Their paper does not state an explicit total running-time bound, but
a straightforward implementation takes $m^{O(n)}$ time.
Our algorithm runs in time polynomial in both $n$ and $m$ in the exact
real-arithmetic model.

\subsection{Proof overview}

\paragraph{Sparse fractional consensus.}
The algorithm maintains a small set $H$ of goods, called the
\emph{core}, to be allocated individually. In each round, we divide
the remaining goods $R=M\setminus H$ into $n$ bundles, temporarily
allowing goods to be split between bundles. We require every agent
$i$ to value each bundle at exactly $v_i(R)/n$. Such a division is
called a \emph{fractional consensus allocation}. We compute a
\emph{sparse} allocation of this kind, in which at most $n(n-1)$ goods
are split between bundles~\cite{GHIMS22}.

\suppressfloats[t]
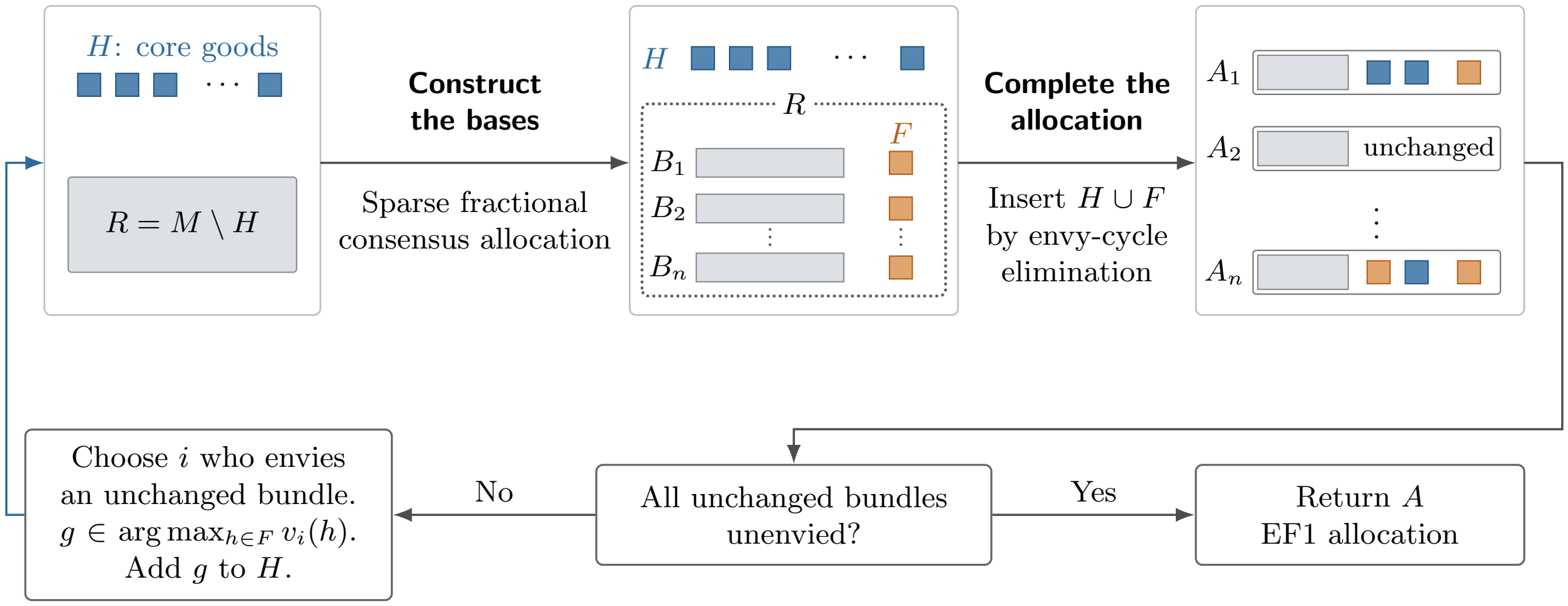
\begin{figure}[t]
\centering
\definecolor{overviewcore}{RGB}{43,105,158}
\definecolor{overviewexception}{RGB}{207,116,34}
\definecolor{overviewbase}{RGB}{221,225,230}
\begin{tikzpicture}[
  font=\footnotesize,
  flow/.style={-{Latex[length=2mm]},line width=0.65pt,draw=black!70},
  panel/.style={draw=black!25,rounded corners=2pt,line width=0.5pt},
  transition/.style={align=center,text width=2.35cm,inner sep=0pt},
  operation/.style={transition,font=\sffamily\footnotesize\bfseries},
  base/.style={draw=black!45,fill=overviewbase,line width=0.4pt},
  core good/.style={draw=overviewcore!85!black,fill=overviewcore!80,line width=0.4pt},
  exceptional good/.style={draw=overviewexception!85!black,fill=overviewexception!65,line width=0.4pt},
  action box/.style={draw=black!60,rounded corners=2pt,fill=white,
               align=center,inner sep=5pt,line width=0.6pt}
]
\draw[panel] (0,0.65) rectangle (2.90,3.90);
\draw[panel] (6.15,0.65) rectangle (9.60,3.90);
\draw[panel] (12.10,0.65) rectangle (15.55,3.90);
\draw[flow] (2.90,2.25) -- (6.15,2.25);
\node[operation] at (4.525,2.90) {Construct\\the bases};
\node[transition,text width=3.10cm] at (4.525,1.65)
  {Sparse fractional\\consensus allocation};
\draw[flow] (9.60,2.25) -- (12.10,2.25);
\node[operation] at (10.85,2.90) {Complete the\\allocation};
\node[transition] at (10.85,1.50)
  {Insert $H\cup F$\\by envy-cycle\\elimination};

\node[text=overviewcore] at (1.45,3.45) {$H$: core goods};
\foreach \x in {0.35,0.75,1.15,2.25}
  \path[core good] (\x,2.95) rectangle ++(0.24,0.24);
\node at (1.90,3.07) {$\cdots$};
\draw[base,rounded corners=1pt] (0.25,1.10) rectangle (2.65,2.10);
\node at (1.45,1.60) {$R=M\setminus H$};

\node[text=overviewcore] at (6.42,3.35) {$H$};
\foreach \x in {6.80,7.20,7.60,9.00}
  \path[core good] (\x,3.23) rectangle ++(0.24,0.24);
\node at (8.50,3.35) {$\cdots$};
\draw[draw=black!65,densely dotted,rounded corners=2pt,line width=0.8pt]
  (6.28,0.85) rectangle (9.47,2.87);
\node[fill=white,inner sep=2pt] at (7.875,2.87) {$R$};
\foreach \y/\j in {2.25/1,1.77/2,1.15/n} {
  \node at (6.56,\y) {$B_{\j}$};
  \draw[base] (6.85,\y-0.15) rectangle (8.40,\y+0.15);
}
\node[text=overviewexception!85!black] at (9.00,2.56) {$F$};
\foreach \y in {2.13,1.65,1.03}
  \path[exceptional good] (8.88,\y) rectangle ++(0.24,0.24);
\foreach \x in {7.625,9.00}
  \foreach \y in {1.39,1.47,1.55}
    \fill[black!70] (\x,\y) circle[radius=0.014];

\foreach \y/\j in {3.20/1,2.40/2,1.10/n} {
  \node at (12.40,\y) {$A_{\j}$};
  \draw[draw=black!55,rounded corners=1pt]
    (12.70,\y-0.23) rectangle (15.30,\y+0.23);
  \draw[base] (12.75,\y-0.18) rectangle (13.70,\y+0.18);
}
\foreach \x/\y in {13.90/3.08,14.30/3.08,14.30/0.98}
  \path[core good] (\x,\y) rectangle ++(0.24,0.24);
\foreach \x/\y in {14.85/3.08,13.90/0.98,14.85/0.98}
  \path[exceptional good] (\x,\y) rectangle ++(0.24,0.24);
\node[font=\scriptsize] at (14.55,2.40) {unchanged};
\node at (14.00,1.72) {$\vdots$};

\node[action box,text width=3.80cm,minimum height=1.05cm]
  (test) at (7.875,-1.45) {All unchanged bundles\\unenvied?};
\node[action box,text width=3.50cm,minimum height=1.65cm]
  (promote) at (1.725,-1.45)
  {Choose $i$ who envies\\
   an unchanged bundle.\\
   $g\in\arg\max_{h\in F}v_i(h)$.\\
   Add $g$ to $H$.};
\node[action box,text width=3.10cm,minimum height=1.05cm]
  (output) at (13.825,-1.45) {Return $A$\\EF1 allocation};
\draw[flow] (15.55,2.25) -- (15.95,2.25)
  -- (15.95,-0.55) -- (7.875,-0.55) -- (test.north);
\draw[flow] (test.east) -- node[above] {Yes} (output.west);
\draw[flow] (test.west) -- node[above] {No} (promote.east);
\draw[flow,draw=overviewcore] (promote.west) -- (-0.40,-1.45)
  -- (-0.40,2.25) -- (0,2.25);
\end{tikzpicture}
\caption{One round of the algorithm.}
\label{fig:value-overview}
\end{figure}

\paragraph{Refining blocks with few queries.}
To compute this allocation with few value queries, we group goods into
blocks, starting with the single block $R$ shared equally among all
$n$ bundles. Linear dependencies among the blocks' value vectors let
us adjust their shares while preserving consensus until at most
$n(n-1)$ blocks remain fractional. We fix the blocks assigned entirely
to one bundle and split each remaining nonsingleton block into two parts of nearly equal
cardinality. Only the new blocks require value queries. Repeating
this procedure leaves only singleton fractional blocks after
$O(\log m)$ stages. This adaptation of consensus-splitting
methods~\cite{AG21,GHIMS22} uses $O(n^3\log m)$ value queries per round.

\paragraph{Setting aside fractional goods.}
Let $B_j$ contain the goods assigned entirely to bundle $j$, and let
$F$ contain the fractional goods. Then $B_1,\ldots,B_n,F$ partition
$R$, with $|F|\le n(n-1)$. Removing the fractional goods gives the
key inequality
\begin{equation}
 \label{eq:overview-base-bound}
 v_i(B_j)\le v_i(R)/n\qquad(i,j\in[n]).
\end{equation}
We query each agent's value for every base and every individual good
in $F$. The values of the core goods in $H$ are already known.

\paragraph{Certifying the completed bundles.}
Starting from the bases, we insert the goods of $H\cup F$ by envy-cycle
elimination. Each insertion goes to an unenvied bundle, while every
agent's own value never decreases. Consequently, bundles receiving new
goods satisfy the EF1 inequalities after deleting their last inserted good.
Only bundles receiving no new good need a final test. If all such
bundles are unenvied, the allocation is EF1.

\paragraph{A failed test grows the core.}
Let $T_i(H)$ be the value to agent $i$ of the core after deleting its
$n$ most valuable goods, or all its goods if $|H|<n$. If $i$ envies an
unchanged bundle, summing the base bounds~\eqref{eq:overview-base-bound}
and the last-insertion inequalities shows that some $g\in F$ satisfies
$v_i(g)>T_i(H)/n$. We add such a good
to $H$ and charge the addition to $i$.

\paragraph{Bounding the number of rounds.}
Every $2n$ additions charged to an agent at least double her trimmed core value
$T_i(H)$.
A short initialization ensures
$T_i(H_0)\ge v_i(R_0)/(2m)$, where $R_0=M\setminus H_0$.
If agent $i$ is charged an addition, she envies a bundle contained in $R_0$,
so $v_i(R_0)>0$ and hence $T_i(H_0)>0$.
Since later additions all come from $R_0$, this value can grow by at
most a factor of $2m+1$. Thus each agent receiving a charge can undergo
only $O(\log m)$ such doublings and can be charged only $O(n\log m)$
additions. Every unsuccessful round adds one good and charges it to one
of the $n$ agents. There are therefore at most $O(n^2\log m)$
unsuccessful rounds.

\subsection{Related work}

Oh, Procaccia, and Suksompong~\cite{OPS21} study the value-query
complexity of fair allocation. In addition to their logarithmic bounds
for up to three additive agents, they give an $O(\log^2 m)$ bound for
three monotone agents and an $O(nm)$ implementation of envy-cycle
elimination for arbitrary monotone valuations.

Our fractional construction draws on consensus splitting.
Goldberg et al.~\cite{GHIMS22} use linear dependencies among value
vectors to leave few fractional goods, and Alon and Graur~\cite{AG21}
refine fractional pieces in necklace splitting. We adapt these ideas
to aggregate queries and combine them with envy-cycle completion.
Feige~\cite{Feige25} studies communication and description complexity
for fair allocation, which measure different resources from the value
queries considered here.

\section{Preliminaries}
\label{sec:preliminaries}

\subsection{Allocation}

There are $n$ agents, indexed by $[n]=\{1,\ldots,n\}$, and a set
$M$ of $m$ goods. Each agent $i$ has a nonnegative additive
valuation $v_i:2^M\to\mathbb R_{\ge0}$:
\[
 v_i(S)=\sum_{g\in S}v_i(g)\qquad(S\subseteq M),
\]
where $v_i(g)=v_i(\{g\})$. A \emph{bundle} is a subset of $M$.

A partial allocation $A=(A_1,\ldots,A_n)$ consists of pairwise
disjoint bundles, where agent $i$ receives $A_i$, and is
\emph{complete} if their union is $M$.

A \emph{fractional allocation} allows each good to be split among
the $n$ bundles, with each bundle receiving a share of the good.
Formally, a fractional allocation of $S\subseteq M$ assigns shares
$x_{jg}\ge0$ with $\sum_{j=1}^n x_{jg}=1$ for every $g\in S$.
Writing $x_j=(x_{jg})_{g\in S}$ for the $j$th fractional bundle,
we extend the valuations to fractional bundles linearly:
\[
 v_i(x_j)=\sum_{g\in S}x_{jg}v_i(g).
\]
A good is \emph{integral} if it belongs entirely to one bundle,
that is, one share is $1$. It is \emph{strictly fractional} if
it is split between at least two bundles.

A \emph{consensus allocation} is a fractional allocation in which
every agent values all $n$ bundles equally. Thus every bundle is
worth exactly one $n$th of the total value to each agent:
\[
 v_i(x_j)=v_i(S)/n\qquad(i,j\in[n]).
\]
Giving each bundle a $1/n$ share of every good always yields
consensus.

\subsection{Fairness}

Given a partial allocation $(A_1,\ldots,A_n)$, agent $i$
\emph{envies} $A_j$ if $v_i(A_i)<v_i(A_j)$.
A bundle is \emph{unenvied} if no agent envies it.
An allocation is \emph{envy-free up to one good} (EF1) if, for
every $i,j$ with $A_j\ne\varnothing$, some $g\in A_j$ satisfies
$v_i(A_i)\ge v_i(A_j\setminus\{g\})$.

If $n=1$, allocating all goods to the single agent is EF1.
If $m\le n$, allocating at most one good to each agent is EF1.
These cases require no queries, so we assume $n\ge2$ and $m>n$.

\subsection{Query model}

A \emph{value query} specifies an agent $i$ and a bundle $S$
and returns $v_i(S)$ exactly.
We count oracle calls across all agents. Running time is measured in
the exact real-arithmetic model: each operation $+,-,\times,\div$
and each comparison of real numbers costs one step. Bundles are
represented explicitly. We include query preparation and output,
but exclude the oracle's own evaluation time.
All logarithms are base two.

\subsection{Envy Cycle}

The \emph{envy graph} of a partial allocation $A$ has one vertex
for each agent and an edge $i\to j$ whenever $i$ envies $A_j$.
Starting from a partial allocation, the envy-cycle elimination
algorithm of Lipton et al.~\cite{LMMS04} repeats the following
steps until all goods are allocated.
\begin{enumerate}
 \item \emph{Eliminate envy cycles.} If the envy graph contains
       a directed cycle $i_1\to i_2\to\cdots\to i_k\to i_1$,
       simultaneously give each agent $i_t$ the bundle of $i_{t+1}$,
       with indices taken modulo $k$. Update the envy graph and
       repeat until it is acyclic.
 \item \emph{Insert a good.} Choose an agent with no incoming
       edge in the resulting graph and add any unallocated good
       to her bundle. Such an agent exists because the graph is
       acyclic, and her bundle is unenvied before the insertion.
       Update the envy graph.
\end{enumerate}
Each cycle exchange strictly increases the value of every
participant's own bundle and leaves all other agents' bundles
unchanged. An insertion cannot decrease the recipient's value,
since valuations are nonnegative. Thus every agent's value for
her own bundle is nondecreasing throughout the algorithm.

To track insertions when bundles change owners, give each initial
bundle a distinct fixed label. A label follows its bundle through
cycle exchanges and is retained when a good is added. Call a bundle
\emph{unchanged} if it receives no new good during completion,
even if it changes owners.

\begin{lemma}[Number of cycle exchanges]
\label{lem:cycle-exchanges}
Between consecutive insertions, and before the first insertion,
envy-cycle elimination performs at most $n(n-1)/2$ cycle exchanges.
\end{lemma}

\begin{proof}
Without insertions, the contents of the $n$ bundles remain fixed.
For each agent $i$, define the rank of a bundle as the number of
bundles she values strictly less. Each rank lies between $0$ and
$n-1$, so the sum of the agents' ranks for their own bundles lies
between $0$ and $n(n-1)$. A cycle exchange gives each participant
a strictly more valuable bundle, increasing her rank by at least
one. Every cycle has at least two participants, so each exchange
increases the rank sum by at least two. The claimed bound follows.
\end{proof}

\begin{lemma}[Last-insertion property]
\label{lem:last-insertion}
Suppose a partial allocation is completed by inserting goods into
unenvied bundles and exchanging bundles, without decreasing any
agent's value for her own bundle. In the final allocation $A$, every
bundle $P$ that received a new good satisfies
\[
 v_i(A_i)\ge v_i(P\setminus\{g\})\qquad(i\in[n]),
\]
where $g$ is the last good inserted into $P$.
If every nonempty unchanged bundle is unenvied, the completed
allocation is EF1.
\end{lemma}

\begin{proof}
Since $g$ is the last good inserted into $P$, the bundle's contents
just before this insertion were $P\setminus\{g\}$. At that time,
the bundle was unenvied, so every agent valued her own bundle at
least as much as $P\setminus\{g\}$. Agents' own values never
decrease, which proves the inequality at completion. If a nonempty
unchanged bundle is unenvied at completion, removing any of its
goods also gives the EF1 inequalities by nonnegativity.
\end{proof}

Thus, when completing an arbitrary partial allocation, the only
possible obstruction to EF1 is an unchanged bundle: some agent may
still envy it after the removal of any one of its goods. Every bundle
that receives a new good satisfies EF1 by the last-insertion property,
regardless of who owns it at completion. In particular, starting
from empty bundles gives a complete EF1 allocation, since every
nonempty final bundle has received a good.

\section{Algorithm and analysis}
\label{sec:values}

The algorithm maintains a \emph{core} $H\subseteq M$ whose
singleton values are known for every agent. Each round constructs
a sparse consensus allocation of $R=M\setminus H$, sets aside its
strictly fractional goods, and uses the remaining bundles as bases.
We complete these bases by inserting the goods set aside together
with the core goods. A failed completion test identifies a good to
add to $H$. We describe these steps in order, then choose the initial
core and bound the number of rounds.

\subsection{Constructing a sparse consensus allocation}
\label{sec:value-blocks}

Equal shares give a consensus allocation, but make every good
strictly fractional. We reduce the number of such goods by treating
groups of goods together. Whenever a group is assigned entirely to
one bundle, its allocation is fixed. Only groups that remain
strictly fractional are divided into smaller groups.

\begin{lemma}[Sparse consensus allocation]
\label{lem:blocks}
For any $S\subseteq M$, $O(n^3\log m)$ value queries suffice to find
a consensus allocation $x=(x_1,\ldots,x_n)$ of $S$ with at most
$n(n-1)$ strictly fractional goods.
\end{lemma}

\begin{proof}
The claim is immediate if $S=\varnothing$, so assume $S$ is nonempty.
All choices in the construction follow fixed deterministic rules.

\paragraph{Working with blocks.}
Throughout the construction, we represent $x$ by a partition of
$S$ into \emph{blocks}. All goods in a block $C$ have identical
shares, and $x_{jC}$ denotes the common value of $x_{jg}$ for
$g\in C$. The contribution of $C$ to $v_i(x_j)$ is therefore
$x_{jC}v_i(C)$. Knowing the block values $v_i(C)$ lets us compute
how changing these shares affects every agent's value for every
bundle. We call a block integral or strictly fractional according
to its goods. Once a block becomes integral, we keep its assignment
fixed and never split it again.

Initially, query $v_i(S)$ for every agent, use the single block
$S$, and set $x_{jg}=1/n$. This is a consensus allocation with
one strictly fractional block. We will maintain consensus and
keep at most $n(n-1)$ blocks strictly fractional at the end of
each stage.

\paragraph{One refinement stage.}
As long as some strictly fractional block contains at least two
goods, perform the following two steps.
\begin{enumerate}
 \item \emph{Split the blocks.} Divide each nonsingleton strictly
       fractional block $C$ into two parts $C_1,C_2$ whose
       cardinalities differ by at most one. Query both $v_i(C_1)$
       and $v_i(C_2)$ for every agent. Give both children their parent's shares:
       $x_{jC_1}=x_{jC_2}=x_{jC}$. No good's share changes, so
       consensus is preserved. Keep singleton blocks as they are.
 \item \emph{Adjust the shares.} Keeping integral blocks fixed,
       change the shares of the strictly fractional blocks until
       at most $n(n-1)$ of them remain. The argument below shows
       how to do this while preserving consensus, using only the
       block values already known.
\end{enumerate}
Splitting makes the blocks smaller but can double their number.
Adjusting the shares restores the bound on their number.
We stop after the adjustment step once every strictly fractional
block is a singleton.

\paragraph{Why the shares can be adjusted.}
Suppose there are $r>n(n-1)$ strictly fractional blocks. In each
such block $C$, choose two positive shares and introduce one signed
transfer amount $z_C$ between them: add $z_C$ to one share and
subtract it from the other. Thus each good's shares still sum to
one. The change in $v_i(x_j)$ is a linear function of the $r$
transfer amounts, with coefficients given by the known block values.
Requiring this change to be zero for every $i\in[n]$ and every
$j<n$ gives $n(n-1)$ homogeneous linear equations in $r$ variables.
Since $r>n(n-1)$, there is a nonzero solution. The value of the
last bundle is also preserved for every agent, because the sum
of her values for all bundles remains $v_i(S)$.

Scale all transfers in this solution by the same positive factor,
increasing it until a decreasing share first reaches zero.
Such a factor exists: all selected shares are initially positive,
and every nonzero transfer decreases one of them. This step
preserves consensus and nonnegativity, eliminates at least one
positive share, and never makes a zero share positive. While more
than $n(n-1)$ blocks remain strictly fractional, repeat this operation,
choosing two positive shares anew in each such block. Each step
reduces the total number of positive shares, so after finitely many
steps at most $n(n-1)$ blocks remain strictly fractional. No value
queries are needed for these computations.

\paragraph{Termination and query count.}
After $k$ refinement stages, every strictly fractional block has
at most $\lceil|S|/2^k\rceil$ goods. Indeed, every nonsingleton
such block is halved in each stage, and integral blocks are never
split again. Hence after at most $\lceil\log|S|\rceil$ stages,
the remaining strictly fractional blocks are singletons. The
adjustment step leaves at most $n(n-1)$ of them, giving the
required bound on strictly fractional goods.

Each stage starts with at most $n(n-1)$ strictly fractional blocks
and uses $2n$ queries for each block that is split. Including the
initial $n$ queries, the total number of value queries is at most
\[
 n+2n\cdot n(n-1)\lceil\log|S|\rceil=O(n^3\log m).
\]
\end{proof}

\subsection{Constructing the bases}
\label{sec:value-bases}

Fix a core $H$ and write $R=M\setminus H$. Apply
Lemma~\ref{lem:blocks} to obtain a consensus allocation $x$ of $R$.
For each $j\in[n]$, the \emph{base}
\[
 B_j=\{g\in R:x_{jg}=1\}
\]
contains the goods assigned entirely to bundle $j$. Let $F$ be the
set of strictly fractional goods in $x$. Then $B_1,\ldots,B_n,F$
partition $R$, with $|F|\le n(n-1)$.

We query every agent's value for every base $B_j$ and every good
in $F$. There are $n$ bases and at most $n(n-1)$ goods in $F$,
so this uses $O(n^3)$ value queries.

The consensus equality gives
\[
 v_i(B_j)+\sum_{g\in F}x_{jg}v_i(g)=v_i(R)/n
 \qquad(i,j\in[n]).
\]
Removing the nonnegative contribution of the fractional goods yields
\begin{equation}
 v_i(B_j)\le v_i(R)/n\qquad(i,j\in[n]).
 \label{eq:cap}
\end{equation}
This key inequality will let us prove progress after a failed
completion test by identifying a sufficiently valuable good in $F$
to add to the core.

\subsection{Completing the allocation}
\label{sec:value-completion}

Starting from the bases constructed in Section~\ref{sec:value-bases},
complete $(B_1,\ldots,B_n)$ by inserting $H\cup F$ using
envy-cycle elimination, and let $A$ be the resulting allocation.
We use the sufficient condition from Lemma~\ref{lem:last-insertion}
to certify EF1: check whether every nonempty unchanged bundle is
unenvied. If this test fails, we use the resulting envy to identify
a good to add to the core.

To measure progress when this test fails, define the
\emph{trimmed core value}
\begin{equation}
 T_i(H)=\min_{\substack{D\subseteq H\\|D|\le n}}v_i(H\setminus D).
 \label{eq:trimmed}
\end{equation}
This deletes the $n$ core goods most valuable to $i$, or all core goods
if $|H|\le n$. The deletion accounts for the at most $n$ last
inserted goods in the completion bounds.
We use $T_i$ only in the analysis; the algorithm does not compute it.

\begin{lemma}[Completion]
\label{lem:completion}
For the completion described above, if all nonempty unchanged bundles
are unenvied, $A$ is EF1.
If agent $i$ envies an unchanged bundle, some $g\in F$ satisfies
$v_i(g)>T_i(H)/n$.
\end{lemma}

\begin{proof}
The first claim follows from Lemma~\ref{lem:last-insertion}.
For the second, the envied unchanged bundle satisfies~\eqref{eq:cap},
so $v_i(A_i)<v_i(R)/n$.
Let $U$ be the set of last inserted goods of the bundles that
received a new good.
These goods are distinct and lie in $H\cup F$.
Since at least one bundle is unchanged, $|U|<n$.

An unchanged bundle has value at most $v_i(R)/n$ by~\eqref{eq:cap}.
For a bundle $P$ that received a new good, let $g$ be its last
inserted good. Additivity and Lemma~\ref{lem:last-insertion} give
\[
 \begin{aligned}
 v_i(P)&=v_i(P\setminus\{g\})+v_i(g)\\
       &\le v_i(A_i)+v_i(g)\le v_i(R)/n+v_i(g).
 \end{aligned}
\]
Thus every final bundle $P$ satisfies
\[
 v_i(P)\le v_i(R)/n+v_i(P\cap U).
\]
The inequality is strict for $P=A_i$, since $v_i(A_i)<v_i(R)/n$.
Summing over the bundles yields
$v_i(M)<v_i(R)+v_i(U)$.
Since $H$ and $R$ partition $M$, subtracting $v_i(R)$ gives
$v_i(H)<v_i(U)$. As $U\subseteq H\cup F$ and $H\cap F=\varnothing$,
additivity lets us write this as
\[
 v_i(H\setminus U)+v_i(U\cap H)
 <v_i(U\cap H)+v_i(U\cap F).
\]
Cancelling $v_i(U\cap H)$ gives $v_i(H\setminus U)<v_i(U\cap F)$.
Moreover, $|U\cap H|<n$, so $U\cap H$ is an allowed deletion set
in the definition of $T_i(H)$. Hence
\[
 T_i(H)\le v_i(H\setminus U)<v_i(U\cap F).
\]
Thus $U\cap F$ is nonempty and contains fewer than $n$ goods,
one of which is worth more than $T_i(H)/n$.
\end{proof}

\subsection{Bounding the number of rounds}

We \emph{charge} each added good to the agent whose envy triggered
its addition. Lemma~\ref{lem:completion} guarantees
$v_i(g)>T_i(H)/n$ for the charged agent $i$. We will show that
every $2n$ additions charged to an agent at least double her trimmed
core value. To bound the number of such doublings, we first choose
an initial core whose trimmed value is at least a $1/(2m)$ fraction
of the remaining value for every agent.

\begin{lemma}[Initial core]
\label{lem:seed}
Using $O(n^2\log m)$ value queries, one can select a core $H_0$
of at most $n(n+1)$ goods such that, for $R_0=M\setminus H_0$,
\begin{equation}
 T_i(H_0)\ge v_i(R_0)/(2m)\qquad(i\in[n]).
 \label{eq:seed}
\end{equation}
\end{lemma}

\begin{proof}
For agent $i$ and a nonempty set $S$, repeatedly split the
current set into two parts of nearly equal cardinality, query
their values, and retain a part of maximum value.
Each step keeps at least half the value. Within
$\lceil\log|S|\rceil$ steps we obtain a good $g$ with
\[
 v_i(g)\ge
 \frac{v_i(S)}{2^{\lceil\log|S|\rceil}}
 \ge \frac{v_i(S)}{2|S|}.
\]
The search uses $O(\log m)$ queries.

Process the agents in order, selecting $n+1$ fresh goods for each,
or all remaining goods if fewer are available. Each search starts
from all goods not yet selected. This uses
$O(n^2\log m)$ queries. If $R_0$ is empty,
\eqref{eq:seed} follows from nonnegativity. Otherwise all selections
were made. Every good selected for $i$ is worth at least
$v_i(R_0)/(2m)$, since its search started from a set containing
$R_0$ with at most $m$ goods. At least one of these $n+1$ goods survives
every deletion of $n$ goods.
\end{proof}

\begin{lemma}[Growth]
\label{lem:growth}
Let $H_0$ satisfy~\eqref{eq:seed} and let $R_0=M\setminus H_0$.
Starting from $H_0$, only add goods outside
the current core. Suppose each added good $g$ is charged to an
agent $i$ and satisfies
\[
 v_i(g)\ge T_i(H)/n,
\]
where $H$ is the core before the addition. Then every agent $i$
with $v_i(R_0)>0$ is charged fewer than
$2n\bigl(1+\lceil\log(2m+1)\rceil\bigr)$ additions.
\end{lemma}

\begin{proof}
Adding goods cannot decrease $T_i$: after any deletion of at most
$n$ goods, the surviving old core is worth at least its previous
trimmed value.

Consider $2n$ additions charged to $i$, starting from a core $C$.
Each new good is worth at least $T_i(C)/n$. Deleting at most $n$
goods leaves at least $n$ of these new goods, worth at least
$T_i(C)$ in total. The surviving part of $C$ is also worth at least
$T_i(C)$. Thus each group of $2n$ additions at least doubles $T_i$.
Intervening additions charged to other agents cannot decrease it.

In any later core, delete the same set of goods that attains
$T_i(H_0)$. All newly added goods lie in $R_0$, so
\[
 T_i(H)\le T_i(H_0)+v_i(R_0)
          \le(2m+1)T_i(H_0),
\]
where the second inequality uses~\eqref{eq:seed}.
If $v_i(R_0)>0$, the seed inequality also gives $T_i(H_0)>0$.
There cannot be $1+\lceil\log(2m+1)\rceil$ doublings, since their
combined factor is greater than $2m+1$. This proves the bound.
\end{proof}

\subsection{Proof of the main theorem}
\label{sec:value-algorithm}

\begin{algorithm}
\label{alg:values}
Construct $H_0$ by Lemma~\ref{lem:seed}, learn every core good's
value to every agent, and set $H=H_0$. Repeat the following round:
\begin{enumerate}
\item Apply Lemma~\ref{lem:blocks} to $R=M\setminus H$,
      obtaining a sparse consensus allocation $x$.
\item Form the bases $B_1,\ldots,B_n$ and the set $F$, and query
      every agent's value for every base and every good in $F$,
      as in Section~\ref{sec:value-bases}.
\item Complete $(B_1,\ldots,B_n)$ by inserting $H\cup F$
      using envy-cycle elimination.
\item If no nonempty unchanged bundle is envied, return the
      allocation. Otherwise choose an agent $i$ who envies one
      and a good $g\in F$ maximizing $v_i(g)$. Add $g$ to $H$
      and charge the addition to $i$.
\end{enumerate}
After a failed round, retain the enlarged core and its singleton
values and recompute the partial allocation of the remainder.
\end{algorithm}

\begin{proof}[Proof of Theorem~\ref{thm:values}]
In Algorithm~\ref{alg:values}, every bundle during completion consists of one $B_j$ and some
goods of $H\cup F$, all with known values. Thus completion,
the final envy test, and the choice of a maximum in $F$
require no queries. Every good added to the core already has
known singleton values.

In a failed round, the selected agent $i$ envies an unchanged
bundle contained in $R_0$, so $v_i(R_0)>0$.
Lemma~\ref{lem:completion} ensures that $F$ contains a good
worth more than $T_i(H)/n$ to $i$. The selected maximum meets
this bound, so the required addition is always possible.
Lemma~\ref{lem:growth} bounds the additions by
$O(n\log m)$ per agent, hence the number of rounds by $O(n^2\log m)$.
At termination, Lemma~\ref{lem:completion} certifies EF1.

Constructing the sparse consensus allocation uses $O(n^3\log m)$
queries per round, and querying the bases and the goods in $F$
uses $O(n^3)$ more. Thus each round uses $O(n^3\log m)$ queries.
Initialization and learning the initial core's singleton values
use $O(n^2\log m+n^3)$ queries. The total is
\[
 O(n^5\log^2 m).
\]

It remains to bound the running time. At each refinement stage, the
$O(n^2)$ fractional blocks have $O(n^3)$ positive shares in total. Each
transfer step makes a positive share zero and never makes a zero
share positive. Thus $O(n^3)$ steps suffice per stage.
Each step uses Gaussian elimination on a system with $O(n^2)$ rows
and columns, then computes the largest transfer scale that keeps
all shares nonnegative. These computations take polynomially many
arithmetic operations.

Lemma~\ref{lem:cycle-exchanges} bounds the number of cycle exchanges
between singleton insertions by $O(n^2)$. Constructing the
envy graph and finding a cycle or an unenvied bundle take polynomial
time. The algorithm has $O(n^2\log m)$ rounds and core goods, at most
$\lceil\log m\rceil$ refinement stages per round, and at most
$|H|+n(n-1)$ insertions
per completion. Thus it performs polynomially many arithmetic and
graph operations. With bundles represented explicitly, all set
operations, query preparation, and output also take polynomial time in $m$.
This proves the claimed running time in $n,m$.
\end{proof}

\section{Discussion}
\label{sec:discussion}

The concurrent work of Lin et al.~\cite{LLTZ26} achieves the optimal
logarithmic dependence on $m$ for EF1. It remains interesting
to combine this dependence with a running-time bound polynomial in
both $n$ and $m$, or with the simple completion argument developed here.
Finally, additivity is used both to express fractional bundle values
in terms of block values and to sum the completion inequalities.
Extending these arguments to broader valuations would require new ideas.

\section*{Declaration of AI use}

The proofs in this paper were found with the assistance of ChatGPT Astra.
The authors subsequently simplified the initial argument and improved
the exposition. Large language models were also used to assist with
writing and editing the manuscript. The authors take full responsibility
for the content of the paper.

\bibliographystyle{plain}
\bibliography{ef1_value_queries_paper}

\end{document}